\documentclass[aps,prb,showpacs,twocolumn,groupedaddress]{revtex4}

\usepackage{algorithm}
\usepackage{algorithmic}
\usepackage[english]{babel}
\usepackage{graphicx}
\usepackage{color}
\usepackage[utf8]{inputenc}
\usepackage{pstricks,pst-grad,color}
\usepackage{graphicx,pifont,amssymb}
\usepackage{amsmath,amssymb}

\def\ket#1{|#1\rangle }

\begin{document}

\title{Spectral functions for single- and multi-Impurity models using DMRG}

\author{Robert Peters}
\email[]{peters@scphys.kyoto-u.ac.jp}
\affiliation{Department of Physics, Kyoto University, Kyoto 606-8502, Japan}

\date{\today}

\begin{abstract}
This article focuses on the calculation of spectral functions for
single- and multi-impurity models using the density matrix
renormalization group (DMRG).
To calculate spectral functions in DMRG, the correction vector method
is presently the most widely used approach. One, however,
always obtains Lorentzian convoluted spectral functions, which in
applications like the dynamical mean-field theory can lead
to wrong results.
In order to overcome this restriction, we 
use chain decompositions in which the resulting
effective Hamiltonian can be diagonalized completely
to calculate a discrete ``peak'' spectrum.
We show that this peak spectrum is a very good
approximation to a deconvolution of the correction vector spectral
function. Calculating this deconvoluted spectrum directly from the
DMRG basis set and operators is the most natural approach, because
it uses only information from the system itself.
Having calculated this excitation spectrum, one can use an arbitrary
broadening to obtain a smooth spectral function, or directly
analyze the excitations.
As a nontrivial test we apply this method to obtain spectral functions
for a model of three coupled Anderson impurities. 
Although, we are focusing in this article on impurity
models, the proposed method for
calculating the peak spectrum can be easily adapted to usual lattice
models.
\end{abstract}


\maketitle

\section{Introduction}
Impurity models can be considered as the most basic models for strongly
correlated electron systems: a small region of interacting electrons
coupled to a reservoir of 
electrons.\cite{hewson1997} The degrees of freedom on
the impurity, the 
degrees of freedom in the reservoir, the coupling between the reservoir and the impurity,
as well as the interactions on the impurity range very widely and depend on
the physical situation of interest.
These situations range from ``real'' impurities in metals, like
cobalt in copper, artificial created nano-structures, to dynamical
mean-field theory (DMFT). 
Besides their wide usage, also the physical effects inherent are very
interesting. The most famous finding is the Kondo-effect
\cite{hewson1997}, which 
manifests itself as a narrow resonance in the single-particle spectrum
of the Anderson impurity model.

To this day, there are many different methods for calculating
properties of quantum 
impurity models: the numerical renormalization group
(NRG)\cite{wilson1975,bulla2008} and
continuous-time quantum Monte Carlo\cite{werner2006,gull2011} are 
two of the currently widely used.  
In this article we focus on the density matrix renormalization
group (DMRG).\cite{white1992,schollwoeck2005,schollwoeck2011}
DMRG has proved to be a highly
accurate method for calculating ground state properties of
one-dimensional models. Thus, it is widely used for calculating
expectation values and correlation functions of such chains gaining
nearly numerical accuracy. DMRG has also already been used 
for calculating impurity 
properties\cite{nishimoto2004,raas2005,raas2005b,nishimoto2006} or as a
DMFT-solver\cite{garcia2004,nishimoto2004b,karski2005,karski2008,raas2009}.

A big advantage of DMRG and NRG is their ability to calculate spectral
functions directly 
in the real frequency domain, making an ill-conditioned analytic
continuation like in Quantum-Monte-Carlo unnecessary.
In NRG and DMRG the Hamiltonian of the impurity model has to be mapped
on a chain model, which is then diagonalized. As there are several
similarities between both methods, we will compare the results of both, 
where it is possible. 

In the next section of this article we introduce the Anderson impurity
model, and we will compare the main features of DMRG and NRG. The
third section is devoted to the calculation of spectral functions for
the single impurity Anderson model. While ground state properties can
be determined with very high precision using DMRG, spectral functions are
much more difficult. We perform our calculations using the
correction vector method,\cite{kuhner1999,jeckelmann2002} resulting in
a Lorentzian broadened spectral function. In the beginning of the
third section we show that the inability of resolving sharp
structures can give wrong results when being used 
in the
self-consistency loop of the DMFT.\cite{georges1996}

The main purpose of the first part of this article is to provide an
extension of the existing correction vector method. We will show how
a peak structure for the spectral function can be calculated by 
using complete diagonalization of an effective Hamiltonian.
Having determined
this peak structure an arbitrary broadening function can be used,
increasing the resolution of the spectral function.
Furthermore, we show
that this peak structure is a 
very good approximation to a deconvolution of the correction vector
spectral function.

Finally, in the last section we will go beyond the single impurity Anderson
model, using the just introduced method for calculating
spectral function for a three impurity model. For testing the newly
developed extension of the correction vector method and the ability of
DMRG to calculate spectral functions for multi-impurity systems, we
couple three Anderson impurity models via a single-particle hopping. 
This will allow in future works to use DMRG as an impurity-solver in
multi-orbital DMFT calculations or as a cluster-DMFT-solver.
\section{Single Impurity Anderson Model}
\begin{figure}[t]
\includegraphics[clip,width=0.8\linewidth]{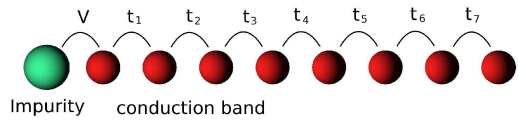}
\caption{(Color online) One-dimensional chain model for the SIAM.}\label{chain}
\end{figure}

We will start with the most simple impurity model, namely the
single impurity Anderson model (SIAM).\cite{hewson1997}
For deriving the Hamiltonian which can be solved via DMRG, we perform a
discretization scheme similar to NRG.\cite{bulla1997,bulla2008}
Following the notation of \textcite{bulla2008}, the Hamiltonian for
the SIAM reads 
\begin{eqnarray}
H&=&\sum_\sigma \epsilon_f f_\sigma^\dagger f_\sigma+Uf_\uparrow^\dagger
f_\uparrow f_\downarrow^\dagger f_\downarrow\nonumber\\
&&+\sum_{k\sigma} \epsilon_k c_{k\sigma}^\dagger
c_{k,\sigma}\nonumber\\
&&+\sum_{k\sigma}V_k\left(f_\sigma^\dagger
c_{k\sigma}+c_{k\sigma}^\dagger f_\sigma\right),
\end{eqnarray}
for which $f^\dagger_\sigma$ represents the impurity and
$c_{k\sigma}^\dagger$ the band states. Herein, $U$ is the amplitude of
a two-particle interaction on the impurity, $\epsilon_f$ the energy level
of the impurity, and $\epsilon_k$ the 
energy dispersion of the conduction band electrons. Finally, $V_k$ represents
the coupling between the impurity and the conduction band.
The hybridization function,
which completely describes the coupling between the impurity and the
bath,\cite{bulla1997} is given by
\begin{displaymath}
\Delta(\omega)=\pi\sum_kV_k^2\delta(\omega-\epsilon_k).
\end{displaymath}
Assuming that the support of the hybridization is covered
by the 
interval $[-1,1]$, we can rewrite the Hamiltonian as
\begin{eqnarray}
H&=&\sum_{\sigma}\int_{-1}^{1} d\epsilon \; g(\epsilon) a_{\epsilon\sigma}^\dagger
a_{\epsilon\sigma}\nonumber\\
&&+\sum_{\sigma}\int_{-1}^1 d\epsilon
h(\epsilon)\left( f_\sigma^\dagger
a_{\epsilon\sigma}+a_{\epsilon\sigma}^\dagger f_\sigma \right)\nonumber\\
&&+\sum_\sigma \epsilon_f f_\sigma^\dagger f_\sigma+Uf_\uparrow^\dagger f_\uparrow f_\downarrow^\dagger f_\downarrow\label{Ham1},
\end{eqnarray}
for which $g(\epsilon)$ and $h(\epsilon)$ represent the dispersion
and the coupling to the impurity, respectively.
The relation between the hybridization function $\Delta(\omega)$,
$g(\epsilon)$ and $h(\epsilon)$ is given by
\begin{displaymath}
\Delta(\omega)=\pi\frac{d g^{-1}(\omega)}{d\omega}h(g^{-1}(\omega))^2.
\end{displaymath}
The Hamiltonian Eq. \ref{Ham1} can now be discretized by dividing
$[-1,1]$ into disjunct intervals $I_n=[k_n,l_n]$ with $\bigcup
I_n=[-1,1]$. In NRG calculations, these intervals should be chosen as
$k_n=\pm\Lambda^{-n}$, $l_n=\pm\Lambda^{-(n+1)}$ with $\Lambda>1$. However,
in DMRG calculations these intervals can be freely chosen. 
As explained in detail in \textcite{bulla2008}, one defines new
fermionic operators for each of these intervals,
so that the Hamiltonian finally becomes that of a one-dimensional
chain, shown in Fig. \ref{chain}, reading 
\begin{eqnarray}
H&=&\sum_\sigma \epsilon_f f_\sigma^\dagger
f_\sigma+Uf_\uparrow^\dagger f_\uparrow f_\downarrow^\dagger
f_\downarrow\nonumber\\
&&+V\sum_\sigma \left(f_\sigma^\dagger c_{0\sigma}+c_{0\sigma}^\dagger f_\sigma\right)\nonumber\\
&&+\sum_{n\sigma}\epsilon_n c_{n\sigma}^\dagger c_{n\sigma}\nonumber\\
&&+\sum_{n\sigma}t_n\left(c_{n\sigma}^\dagger c_{n+1\sigma}+c_{n+1\sigma}^\dagger c_{n\sigma}\right).
\end{eqnarray}

\subsection{NRG versus DMRG calculations}
As we are going to give several NRG results for comparison, let us
shortly describe the differences between NRG and DMRG.
For detailed
information about how these methods work, 
we refer to the reviews by \textcite{schollwoeck2011} (DMRG) and
by \textcite{bulla2008} (NRG).
Recently, there have also been several 
attempts to combine both methods.\cite{saberi2008,weichselbaum2009,holzner2010,pitzorn2011}

The main recipe for NRG is: Iteratively increase the length of the
chain starting at the impurity site, diagonalize the Hamiltonian
and truncate the high energy states,
if the number of Fock space states exceeds a certain number $N$. This
procedure can only work because the intervals during the
discretization were logarithmically arranged leading to an
exponentially decreasing coupling $t_n$ in the one-dimensional chain.
Thus, having diagonalized the chain up to site $i$,
one can assume that the remaining sites of the chain are only a small
``perturbation'', as the coupling $t_i$ is small compared to the energy
difference between the high energy states and the low energy
states. Being only interested in the energetically low lying states, one can
truncate the high energy states and treat them as approximate excited
states of the whole chain, thus obtaining information about the whole 
spectrum of energies.

On the other hand, in DMRG one calculates from the beginning the
ground state of the whole chain. DMRG combines left blocks and
right blocks to calculate approximate ground states of the whole
chain. The left and right blocks are gradually refined by calculating
the ``best'' basis sets for those blocks, which best describe the
ground state of the chain. This can be done by calculating the density
matrix of the ground state for the blocks. Thus, DMRG is able to
calculate very accurate 
ground states for one-dimensional chains, and does not necessarily need
a logarithmic discretization of the conduction band.
 
One big difference between both methods is, that using NRG one can
calculate a complete set of eigenstates of the model.
\cite{peters2006,weichselbaum2007}
This can only be achieved by the above described ``energy separation''
due to the logarithmic discretization. 
The disadvantage is that this logarithmic discretization
reduces the resolution of high energy contributions, as they are worse
resolved during the discretization.
Furthermore, in NRG calculations one has to include always all states
contributing to one ``energy shell'' (imposed by the discretization
parameter $\Lambda$) for
setting up the Hamiltonian up to site $i$. Going from the single
impurity model to multi-impurities, having more than one conduction
band, will lead to an exponential
increase in matrix dimensions, eventually making calculations
impossible or inaccurate with increasing number of conduction bands.
On the other hand, in DMRG calculations one can easily separate each
conduction band, as they only couple directly at the impurity to each
other. Thus, only the impurity itself
must be considered increasingly
difficult within DMRG, as all conduction bands couple at this site.
Additionally, in DMRG the
discretization of the conduction band can be freely chosen, increasing
the resolution of high energy contributions. These are clearly
advantages of the DMRG. The price one pays is loosing
the complete set of excited states calculated within NRG, because DMRG
can only calculate
properties of the ground state or a very few excited states.

\section{Spectral functions}
\subsection{Definition and Calculation}
In this article we focus on the calculation of dynamical properties,
such as Green's functions. As Green's functions depend
on excited states, the calculation of these dynamical properties can be
easily performed within NRG having a complete basis set.
Within DMRG it is much more difficult, because the usual
DMRG-basis set is optimized for the ground state, but excited states
are not optimally represented. 

Nevertheless, there are ways to calculate the Green's function within DMRG.
The definition of the fermionic one-particle Green's function at
$T=0$ corresponding to the ground state $\vert \Psi_0\rangle$,
is given by
\begin{eqnarray}
iG(t)&:=&\Theta(t)\langle\Psi_0\vert [c(t),
  c^\dagger(0)]_+\vert\Psi_0\rangle\nonumber\\
G(\omega)&=&\sum_i\left(\frac{\langle \Psi_0\vert c\vert
  \phi_i\rangle\langle\phi_i\vert
  c^\dagger\vert\Psi_0\rangle}{\omega+E_0-E_i+i\eta}\right.\nonumber\\
&&+\left.\frac{\langle \Psi_0\vert c^\dagger\vert
  \phi_i\rangle\langle\phi_i\vert
  c\vert\Psi_0\rangle}{\omega-E_0+E_i+i\eta}\right)\label{leh}\\
&=&\Bigl\langle \Psi_0\Bigl\vert c\frac{1}{\omega+E_0-H+i\eta}
  c^\dagger\Bigr\vert\Psi_0\Bigr\rangle\nonumber\\
&&+\Bigl\langle \Psi_0\Bigl\vert c^\dagger\frac{1}{\omega-E_0+H+i\eta}
  c\Bigr\vert\Psi_0\Bigr\rangle,\label{cor}
\end{eqnarray}
with $\eta\rightarrow 0$, and $\ket{\phi_i}$ being a complete Fock
space basis of 
eigenstates. Equation \ref{leh} is
commonly called Lehmann representation and is used to calculate
spectral function within NRG.\cite{peters2006,weichselbaum2007}

Unfortunately, such a complete basis set of eigenstates,
$\ket{\phi_i}$, is usually not available
within DMRG, as the emerging matrix sizes are too large for full
diagonalization, so that iterative methods, such as Lanczos or
Jacobi-Davidson, for calculating the ground state are used.
However, as it is possible to calculate the result of an operator
acting on a state, one can use Eq. \ref{cor} for calculating the
Green's function. The operator
$\frac{1}{\omega+E_0-H+i\eta}$ 
acting on $c^\dagger\ket{\Psi_0}$ results in the so called
correction vector,\cite{kuhner1999,jeckelmann2002} from which a
Lorentzian
broadened spectral function can be calculated, corresponding to Eq. \ref{cor}.
The physical interesting spectral function would be obtained by taking
the limit $\eta\rightarrow 0$.
However, this limit $\eta\rightarrow 0$ cannot directly be performed,
because the spectrum of a finite chain is discrete, giving a 
finite number of peaks at $\delta(\omega+E_0-E_i)$. 
Therefore, one would have to know exactly $E_i$ and the corresponding
eigenstate $\vert E_i\rangle$ (being equivalent to the knowledge of
a complete basis set of eigenstates). Having no complete basis
set of eigenstates, it is impossible to numerically calculate the operator
$\frac{1}{\omega+E_0+i\eta-H}$ for $\eta\to 0$.
Decreasing $\eta$ makes it more and more difficult to calculate a
converged correction vector, increasing the necessary truncation
dimension of the basis and the computation time.

Therefore, one has to use a finite $\eta>0$, resulting in a Lorentzian
broadened spectral function. 
There are deconvolution schemes\cite{raas2005b} like maximum entropy, 
calculating a Green's 
function in the limit $\eta\rightarrow 0$ starting from the
broadened correction vector spectral function, but their ability to 
resolve sharp features is rather limited.
Furthermore, when using deconvolution schemes one usually has to allow for
some small discrepancies between the convolution of the result and the
DMRG spectral function.
This introduces a new source of errors and arbitrariness into
the final result. 
Other possibilities for calculating spectral
functions within DMRG are expanding the spectral function
into a continued fraction,\cite{hallberg1995,dargel2011} expanding
in Chebyshev polynomials,\cite{holzner2011} or to use a Fourier
transformation from a real-time calculation.\cite{pereira2009} 
However, to this day the correction vector method is the most widely used
method.

\subsection{Problems within Dynamical Mean-Field Theory}
Before showing a very natural way for obtaining a deconvoluted
spectrum, we want to show the occurring problem when using DMRG in a DMFT
self-consistency calculation. DMFT calculates a solution of a lattice
model, like the Hubbard model, by mapping it onto a self-consistent
impurity calculation.\cite{georges1996} This self-consistency is
usually obtained by iteratively solving the impurity model.
Therefore, one has to deconvolute the
result in every DMFT iteration, so that errors introduced by
convolution/deconvolution can grow during the DMFT cycle.

Especially the frequencies around the Fermi energy are 
most important for the stabilization of a converged DMFT solution.
The results can be entirely different depending on the existence of
a gap at the Fermi energy.
\begin{figure}[t]
\includegraphics[clip,width=0.8\linewidth]{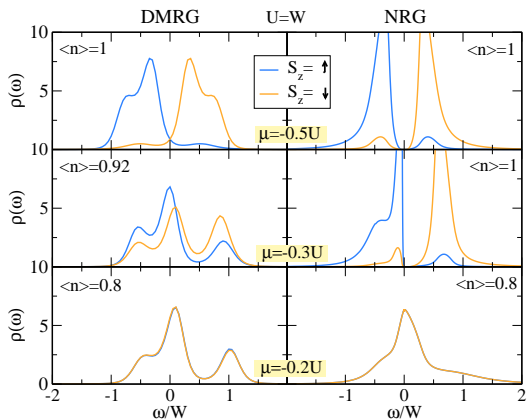}
\caption{(Color online) 
Comparison between spectral functions, $\rho(\omega)=- \mathfrak{Im}(G(\omega))/\pi$, of DMFT calculations for the
  antiferromagnetic N\'eel state in the Hubbard
  model (Bethe lattice, $U=W$, $T=0$, different chemical potentials
  $\mu$) using DMRG and NRG as impurity solver. The left 
  panels are calculated via DMRG, the right panels via NRG. Blue and
  orange lines correspond to spin-up and spin-down, respectively.
For the DMRG correction vector we used $\eta=0.25 W$.}\label{dmft}
\end{figure}
An example for this is given in Fig. \ref{dmft}. The figure shows DMFT
results for the antiferromagnetic state in the Hubbard model,
\begin{displaymath}
H=t\sum_{\langle i,j\rangle,\sigma} c_{i\sigma}^\dagger c_{j\sigma}+U\sum_in_{i\uparrow}n_{i\downarrow},
\end{displaymath}
for $U=W$, $U$ being the local interaction strength and $W=4t$ the bandwidth
of the non-interacting electron system for a Bethe lattice.
The results as given by NRG show a phase separation
between the antiferromagnetic phase at half filling and the
paramagnetic state away from half filling.\cite{peters2009b} The shown
DMRG results are deconvoluted 
by a non-biased 
Maximum Entropy scheme.\cite{raas2005b} Non-biased means that we have not
used any additional assumptions 
for the result of the spectral function. For the DMRG
correction vector, a broadening of $\eta=0.25W$ has been used.  
This example is somewhat extreme using a very large broadening.
However, as all structures in the spectral function must be included
into the discretization interval $[-1,1]$, such situations can occur, as the
position of structures depend not only on the bandwidth $W$, but
also 
on the interaction strength and the chemical potential.
In this specific chosen example the features in the original
convoluted spectral 
function as calculated by DMRG are smeared out.
Deconvolution is able to sharpen the contours, but it
cannot resolve detailed structure. 

The half filled solutions of DMRG and NRG in Fig \ref{dmft}
look quite similar. The main difference is that there
is no real gap at the Fermi energy, $\omega=0$, in the DMRG result.
The situation is worse for $\mu=-0.3W$, corresponding to a
non-particle-hole-symmetric situation. The NRG/DMFT result is
still an antiferromagnetic half filled N\'eel state, $\langle
n_\uparrow+ n_\downarrow\rangle=1$. This state is still
gapped at the Fermi-energy, but the lower Hubbard band has moved
towards the Fermi-energy. DMRG cannot resolve this very sharp
structure near the Fermi-energy, and entirely closes the gap during
the self-consistency cycle, even though deconvolution has been used. Thus,
it appears, as if the DOS for the spin-up and spin-down components are
just shifted 
apart from each other, resulting in a doped
antiferromagnetic metallic state. We want to emphasize, that this
DMRG/DMFT result is not just a differently broadened version of the
NRG/DMFT result, but is a different solution due to the
self-consistency.
Only again for larger doping NRG and DMRG show both a paramagnetic
metallic state, which reasonably agree with each other.

\subsection{Calculating the peak spectrum}
\begin{figure}[htp]
\includegraphics[width=0.8\linewidth]{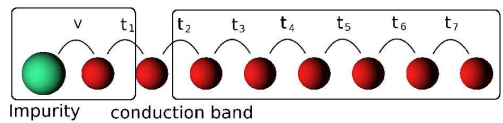}
\caption{(Color online) DMRG chain with impurity, showing
    a position at which the effective Fock
  space dimension is reduced due to the open boundary.}\label{chain2}
\end{figure}

In general it is impossible to obtain a complete basis of
  eigenstates for the whole chain, because the Fock space dimension is too
  large. However, DMRG 
  automatically creates basis sets for parts of the chain best
  describing the aimed states, e.g. the ground state. From these basis
  sets an effective Hamiltonian is set up and diagonalized.
Usually working with high truncation dimensions to obtain an accurate
result, the resulting dimensions for the effective Hamiltonian are
still too large for complete diagonalization. (By
  ``complete diagonalization'' we mean, that all eigenstates and
  eigenvalues are calculated.)
 Nevertheless, there are several possibilities to obtain an
  effective description of the chain consisting of a Fock space
  dimension, which is small enough for complete diagonalization: e.g. left
  and right block can be strongly truncated and combined without a
  single block in between, or at the open boundaries of the chain, at
  which the basis dimension is in general smaller due to the boundary,
  see Fig. \ref{chain2}.
\begin{figure}[htp]
\includegraphics[width=0.8\linewidth]{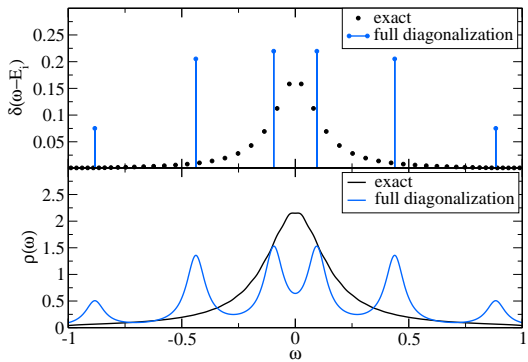}
\caption{(Color online) Impurity spectral functions calculated by a complete
  diagonalization of an effective Hamiltonian. The chain decomposition consists of two
    local sites on the left side and a truncated ``right-block'' (linear discretization of the
  conduction electrons, $U=0$, $\Delta=0.1$, DMRG-truncation $m=300$). 
Upper panel: spectral weights from DMRG compared to the exact result
calculated by diagonalization of the non-interacting hopping chain.
Lower panel: Lorentzian broadening of the spectral weights using $\eta=0.05$.
}\label{without}
\end{figure}
Having a small effective basis set describing the whole
chain, one can 
completely diagonalize the effective Hamiltonian and 
calculate the discrete peaks of the spectral function.

Figure \ref{without} shows the resulting peaks in the spectral
function, $\rho(\omega)=-\mathfrak{Im}(G(\omega))/\pi$, and a Lorentzian broadening of these peaks for the
non-interacting SIAM, $U=0$, compared to the exact result calculated
from the
discretized one-particle Hamiltonian.
First of all, it should be clear that the
number of calculated peaks is small, and the position and weight of
the peaks do not agree with the
exact result. Although there are only $6$
peaks visible, the used Fock space basis consists of 
approximately $100$ states at this point. However, most of the excited states
have zero weight in the spectral function.
Thus, this straightforward implementation of 
calculating the spectral function by 
a complete diagonalization of an effective Hamiltonian
yields not enough contributing states and fails. This corresponds to the
optimization of the basis towards the ground state of the system. What
is needed, is a change in the basis set towards excited states
contributing to the spectral weight.
This optimization of the
DMRG basis can be achieved by the already introduced correction vector
\cite{kuhner1999,jeckelmann2002} 
\begin{eqnarray}
&&\frac{1}{\Omega+E_0-H+i\eta}c^\dagger\vert\Psi_0\rangle\nonumber\\
&=&\sum_i\vert E_i\rangle\langle E_i\Bigl\vert \frac{1}{\Omega+E_0-E_i+i\eta}c^\dagger\Bigr\vert\Psi_0\rangle\nonumber.
\end{eqnarray}
We label the frequency at which the correction vector is
calculated as $\Omega$.
Obviously, excited states, for which $\Omega+E_0\approx E_i$ holds, have strong
weight in the correction vector.
Thus,
trying to optimize the ground state and the correction vector will
optimize the basis towards contributing states. Of course, this is not
surprising, as the correction vector exactly describes the spectral
function.\cite{kuhner1999,jeckelmann2002}  However, instead of
calculating only the value of the spectral function at $\Omega$, we
now completely diagonalize the set up Hamiltonian, which  yields the
possibility to directly  
observe the contributing excited states $\vert E_i\rangle$. 
A similar idea has already been used in
\textcite{kuhner1999}, in which the correction vector was used for
adapting the DMRG basis set followed by the calculation of the spectral function
via the Lanczos method.

\begin{figure}[htp]
\includegraphics[width=0.8\linewidth]{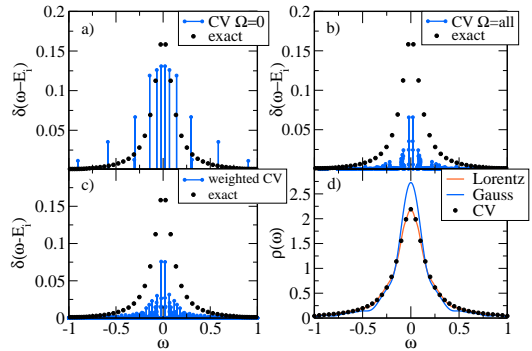}
\caption{(Color online) Spectral weights calculated by diagonalization including the
  correction vector for the non-interacting SIAM $U=0$,
  $\Delta=0.1$. The conduction 
  electrons were discretized linearly using $N=50$ sites. The
  correction vector used $\eta=0.05$. a) spectral weights
of the correction vector (CV) for $\Omega=0$ compared to the exact
results. (Chain decomposition as in Fig. \ref{without}) b) spectral weights of correction vectors
$\Omega=\{-1.55,1.5,...,1.5,1.55\}$. c)
spectral weights of the correction vectors weighted according to
their position. d) Lorentzian ($\eta=0.05$) and Gaussian ($b=0.045$)
broadened spectral 
functions compared to data points of the correction vector.}\label{withU0} 
\end{figure}
\begin{figure*}[t]
\includegraphics[width=0.8\linewidth]{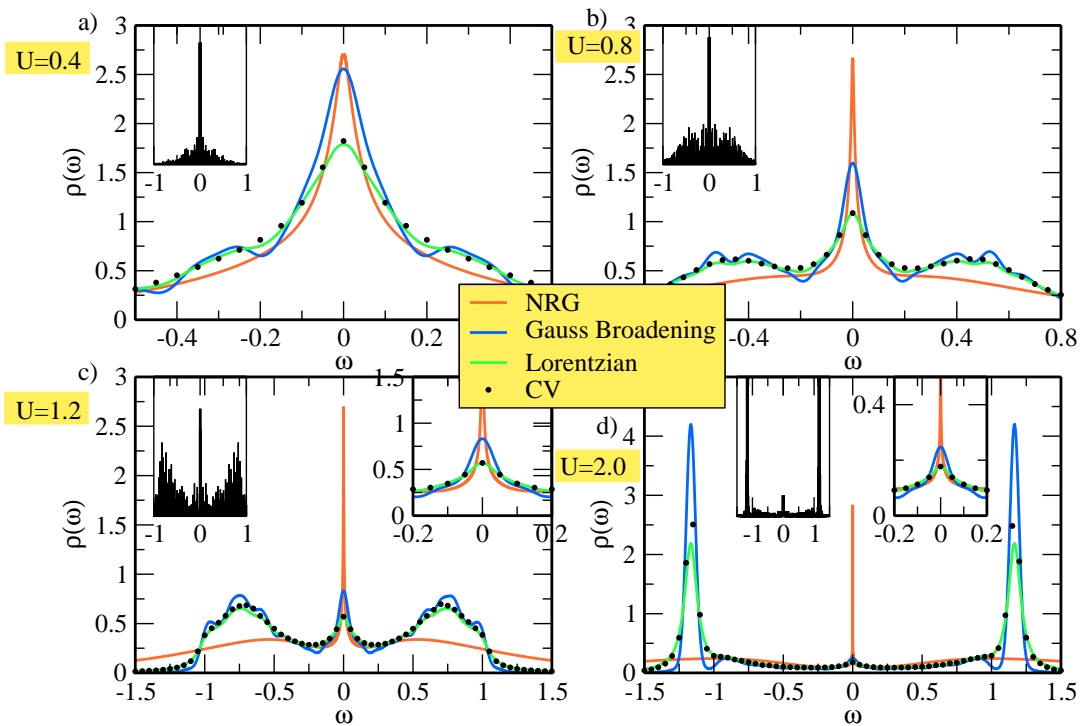}
\caption{(Color online) Spectral function for the SIAM ($\Delta=0.1$, linear
  discretization $N=50$ sites, Chain decomposition as in Fig. \ref{without}) for different interaction strengths $U$
  (upper panel: $U=0.4$ (left); $U=0.8$ (right). lower panel: $U=1.2$
  (left); $U=2.0$ (right). We compare between the usual correction
  vector (CV) result $\eta=0.05$, a Gaussian broadening ($b=0.045$) of the
  calculated peaks, a Lorentzian broadening using $\eta=0.05$, and the NRG-result. The left inset in each panel
  gives the calculated level structure. The right inset in the lower
  panels gives a magnification around the Fermi energy.}\label{withU} 
\end{figure*}
Figure \ref{withU0} shows again the calculated spectral weights using
correction vectors, $\eta=0.05$, for the non-interacting SIAM, $U=0$, and
$\Delta=0.1$. For comparison the exact peaks are included again.
It should be stated that one cannot expect 
that DMRG calculates the exact positions of these peaks, as DMRG
works in a  
restricted basis setting up an effective Hamiltonian.
However, it is important that the
summed up spectral weight around a frequency is approximately the
same as in the exact result. Panel a) shows the spectral weight using
only the correction vector at $\Omega=0$. Comparing to
Fig. \ref{without}, the weight is 
now distributed over more states, especially near $\omega=0$. But for
$\vert \omega\vert>0.2$ there are only a very few contributing
states. Panel b) shows the spectral weights as a sum of all used
correction vectors, $\Omega=\{-1.55,1.5,...,1.5,1.55\}$. 
For each frequency $\Omega$ a separate correction vector calculation
followed by a complete diagonalization of the effective Hamiltonian is
performed. 
As the final
spectral function should be normalized, the weight of the peaks is
divided by the 
number of used correction vectors. 
Again, the spectrum shows regions
having only very small spectral weight, e.g. $\omega\approx 0.15$,
although the exact result 
shows weight in those regions. The main reason for this
behavior is the renormalization. If only a
few correction vectors are contributing in those
regions, the weight is renormalized, when dividing by the number of
used correction vectors. One can easily overcome this problem
by taking into account the position of the correction vectors,
$\Omega$, as the spectral weights are supposed to be most accurate
around this frequency. Thus, we can assign a weight to the spectral peaks using
a Lorentzian function having the same $\eta$. Denoting a peak at $E_i$
calculated by the correction vector $\Omega$ as $p_i(E_i,\Omega)$, we
transform all peaks to
\begin{displaymath}
P_i(E_i,\Omega)=\frac{1}{Z}\frac{\eta}{\pi}\frac{1}{(\Omega-E_i)^2+\eta^2}\;p_i(E_i,\Omega),
\end{displaymath}
in which $Z$ is a constant factor normalizing the sum of the spectral
weight to unity. The result of this transformation can be seen in panel
c) of Fig. \ref{withU0}. The distribution of the spectral weights
follows now the
exact result. Finally these peaks can be broadened using an arbitrary
broadening-functions. Panel d) shows the result using a Gaussian and a
Lorentzian broadening. For the Lorentzian broadening we use the same
$\eta$ as 
for the correction vector. 
For the Gaussian broadening we use
$\delta(\omega-E)\to\frac{1}{b\sqrt\pi}\exp\left(-(\omega-E)^2/b^2\right)$.
The result of this Lorentzian broadening
is in very good agreement with the correction vector points.
We want to stress this point, as it means that the calculated
peak structure represents a deconvolution of the correction vector
spectral function. This deconvolution is calculated directly within
the DMRG basis. No additional methods or assumptions have to be used.
Using a Lorentzian broadening upon the exact spectral poles,
calculated from the one-particle Hamiltonian,
agrees with this calculated curve.
Unfortunately, it is not the general case that the Lorentzian
broadening of the calculated peaks yields exactly the same result as
the correction vector spectral function.
Usually, there
will be small derivations, due to the changing of the basis set for
every correction vector. Nevertheless, the calculated peak structure
gives a very good approximation to a full deconvolution of the correction
vector result, which can be seen in the next examples.

Figure \ref{withU} shows several results for the interacting SIAM for
$\Delta=0.1$. 
The interaction strengths range from $U=0.4$ to $U=2.0$. We chose a
linear discretization of the conduction band with $N=50$ sites for the
DMRG calculation and compare between the usual correction vector (CV)
result, NRG, and a broadening of the calculated peak structure. 
For $U=0.4$ ($U/\Delta=4$), there is still a broad and
Lorentzian like peak at $\omega=0$.
For this interaction strength, all curves reasonably agree with each
other, taking into account that different broadening functions were used.
Increasing the interaction strength, one can observe
how the typical three-peak structure in the SIAM evolves. These three
peaks can also be nicely observed in the calculated peak
structure itself. Even for $U=2$, for which the Kondo resonance is hardly
visible in the broadened DMRG results, one can still observe a peak
near $\omega=0$ in the peak structure. Obviously, NRG can resolve the
Kondo resonance for those interaction values, but fails to
precisely resolve the Hubbard bands. Nevertheless, using
a discretization parameter $\Lambda=2$ and a logarithmic-Gaussian
broadening ($b=0.8$),\cite{bulla2008} also NRG fails to precisely
obtain $\rho(0)=1/(\Delta\pi)=3.183$. Although there are ways to
improve this result (z-averaging or directly calculating the
self-energy)\cite{bulla2008}, these techniques were not used here to
present a comparison to a ``basic'' NRG-calculation. 
DMRG results show a
homogeneous resolution for the whole spectral function. 
The Lorentzian broadening of the calculated peaks, agrees in all four
examples very well to the correction vector points. Again, this is
equivalent to the statement that the calculated peak structure
corresponds to a deconvolution of the correction vector spectral
function. 

The spectral functions in Fig. \ref{withU} calculated by
broadening of the peaks, show some oscillations, especially in the
Hubbard peaks. These oscillations are stronger pronounced in the
Gaussian broadened spectral functions, because the Lorentzian
broadening smooths those oscillations. The shape of those
structures depend on the number of used correction-vectors, the
number of states kept during the DMRG-sweep 
and the exact decomposition of the chain to create
the effective Hamiltonian. 
\begin{figure}[tb]
\includegraphics[width=0.8\linewidth]{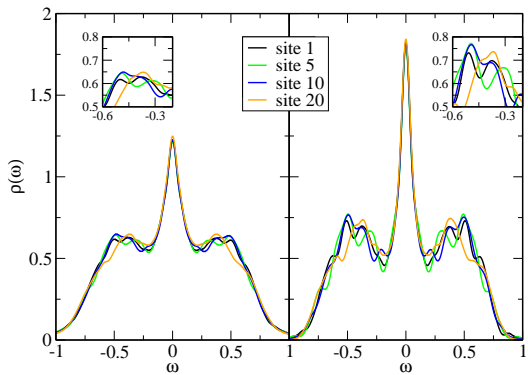}
\caption{(Color online) Spectral function for the SIAM ($\Delta=0.1$,
  $U=0.8$, linear discretization $N=50$ sites). Left: Lorentzian
  broadening. Right: Gaussian broadening ($b=0.045$). A two-block
  decomposition is used for calculating the peak structure. ``Site''
  corresponds to the rightmost site in the left block. The impurity is
located at site 0. The insets show a magnification of the
Hubbard peak.}\label{diffblock}  
\end{figure}
Figure \ref{diffblock} shows the calculated spectral
function created 
by different chain decompositions for $U/\Delta=8$. For all shown
spectral functions a 
two-block decomposition was used, for which the label ``site''
corresponds to the rightmost site in the left block. 
To obtain an effective basis which can be completely
diagonalized the blocks are additionally truncated to $100$ states.
This figure
illustrates the dependence of the oscillations on the decomposition
of the chain. However, this figure also shows that the proposed
algorithm is not limited to decompositions near the open boundary at
which the impurity is located.
When the spectral functions is calculated in a
decomposition, in which the impurity is located in a truncated block,
the spectral weight does not have to sum up to unity. However, when
the correction vector is included into the density matrix as target
state, the spectral weight is reduced in the shown examples by
approximately $10\%$, keeping $m=300$ states during the
calculation. The spectral weight was normalized to
unity in Fig. \ref{diffblock}. 
The oscillations are naturally more pronounced in the Gaussian
broadening, as this broadening is used here to enhance the resolution
of small structures. 

\begin{figure}[tb]
\includegraphics[width=0.8\linewidth]{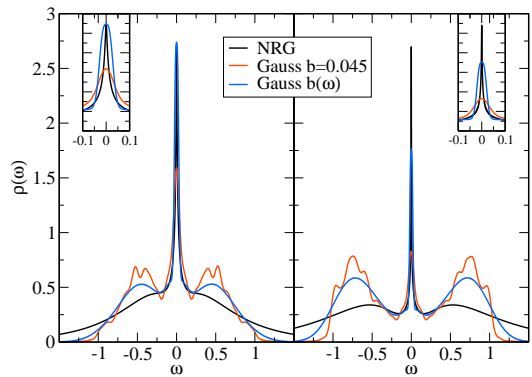}
\caption{(Color online) Spectral functions for the SIAM ($\Delta=0.1$, $N=50$
  sites) using different broadening schemes. Left panel: $U=0.8$ Right
  panel: $U=1.2$. The insets show a magnification around the Fermi
  energy.}\label{diffbroad}  
\end{figure}
One very big advantage having calculated the peak structure is
obviously that one can immediately change the broadening of the
peaks. To resolve the Kondo resonance at $\omega=0$ without
introducing too many oscillations for $\vert \omega\vert>0$ one can now
use a 
frequency dependent broadening focusing on the Kondo resonance. In  
Fig. \ref{diffbroad} we compare the NRG result to an usual Gaussian broadened
spectral functions and a frequency-dependent broadening, in which the
position of the 
peak determines the broadening parameter as $b(E_i):=\vert 1.5 E_i\vert$.
This allows for sharp structures at $\omega\approx 0$, but smoothing
oscillations for large $\omega$. 
The shown results are for $U=0.8$ (left panel) and $U=1.2$ (right panel).
Especially for $U=0.8$, the frequency dependent broadening resolves 
well the Kondo resonance and the Hubbard bands. The calculated shape of the
Kondo-resonance, of course, depends on the used broadening. As NRG
uses a Gaussian-logarithmic broadening,\cite{bulla2008} the shape differs from the DMRG
result. Whereas the NRG result determines a smaller width of the
Kondo-resonance for $U=1.2$, the width of the DMRG result stays nearly
unchanged. 
To understand this, it is advisable to directly analyze the
calculated peak structure.
\begin{figure}[tb]
\includegraphics[width=0.8\linewidth]{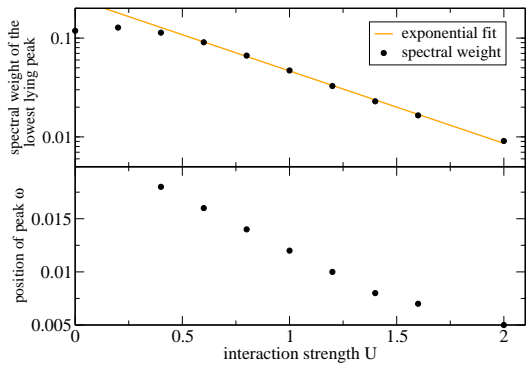}
\caption{(Color online) Analysis of the lowest lying excitation for the SIAM
  $\Delta=0.1$ and $N=50$ sites for different interaction
  strengths. Upper panel: weight of the lowest lying peaks on
  logarithmic scale. Lower panel: position of the lowest lying
  peak.}\label{lowestpeaks}  
\end{figure}
In Fig. \ref{lowestpeaks} we analyze the energetically lowest peaks in
the spectral 
function of the SIAM as seen by the DMRG. The upper and the lower
panel give the weight and the position of the peak, respectively.
For the interaction values under consideration the spectral weight
begins to exponentially decrease, which is the expected behavior for
Kondo-physics. Also the position of the peak moves closer to $\omega=0$
when increasing the interaction strength. As this peak is supposed to
describe the Kondo-resonance, its position is supposed to
exponentially approach $\omega=0$ with increasing interaction strength.
However, this is clearly not the case. The expected Kondo temperature
for $U=2.0$ and $\Delta=0.1$ is 
approximately $T_K\approx 10^{-4}$, but the lowest lying peak is located
at $\omega\approx 0.005$.  
This is because the chain has been discretized linearly 
with $N=50$ sites, setting an energy scale of $\Delta E=0.04$, as the
conduction band electrons have energy from $-1$ to $1$. Structures
below this value are hard to resolve. Unfortunately, increasing the
number of sites or changing the discretization helps only to a limited
amount. The resolution is also limited by the number of Fock space
states, for which the complete diagonalization is performed, by the
truncation of the DMRG basis set, and the used $\eta$ in
the correction vector. We also performed calculations using a
logarithmic discretization of the conduction band. The result is
that for example the position of the lowest peak moves for $U=1.2$
from $\omega\approx 0.01$ to $\omega\approx 0.007$, thus giving only
a little improvement for the Kondo resonance compared to the linear
discretization. However, the high energy features of the spectral
function turned out to be much more oscillating. The best results
for spectral functions were obtained by using a linear
discretization mesh corresponding to $\eta$ in the correction
vector.
Thus, though we can increase the resolution by calculating the peak
structure, the resolution of the spectral function remains limited
due to truncation and discretization of the chain. 

\begin{figure}[tb]
\includegraphics[width=0.8\linewidth]{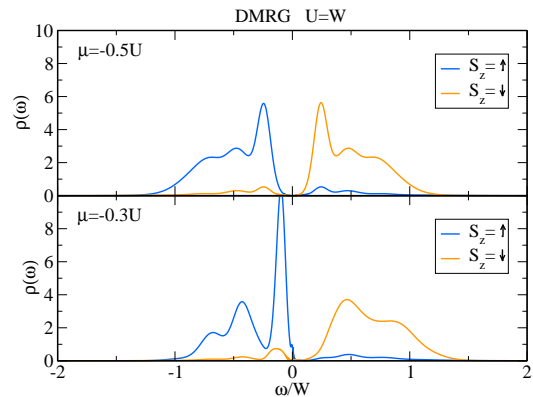}
\caption{(Color online) DMFT calculations (Bethe lattice, $U=W$) using DMRG as impurity
  solver. The spectral functions are calculated using a Gaussian
  broadening upon the calculated peak spectrum with broadening $b=0.045$.
}\label{dmftleh}  
\end{figure}
Finally, let us come back to the problem of convolution/deconvolution
within DMFT calculations. Figure \ref{dmftleh} shows again results for
the antiferromagnetic phase in the Hubbard model for $U=W$ (compare to
Fig. \ref{dmft}). This time the
spectral functions were calculated using the above described method
for determining the peak spectrum and later convoluted using a
Gaussian broadening to obtain smooth curves. These results agree now
very well with the known results from NRG. Both solution correspond to
half filling showing a gap at the Fermi energy. Thus, we were able to
improve the resolution of the spectral functions using the above
described method.

Let us shortly summarize this section. We have shown that one can
gain easily additional information about the spectral function while
doing a correction vector calculation. These information are obtained
by a complete diagonalization of the effective Hamiltonian created by the
DMRG-basis including the correction vector. This is possible because
the Fock space is small close to the open boundary, where the impurity
is located. As this complete diagonalization is only performed, when a
converged DMRG basis set for a correction vector is found, the
additionally used time is negligible. 
Changing the correction vector, also changes the DMRG
basis, finally giving a dense spectrum of peaks. 
This peak structure approximates a full deconvolution of the
correction vector spectral function, in the sense that a Lorentzian
broadening of these peaks results nearly in the correction vector
spectral function. Besides representing a deconvolution, one can
analyze those peaks directly or use an arbitrary broadening function.

\section{Going Beyond Single Impurity Calculations}
Besides being able to calculate spectral functions of the single
impurity Anderson model with homogeneous resolution, DMRG is also able to
perform calculations for multi-impurity systems, for
which the model consists of more than one conduction channel and
possibly more than one interacting sites. In NRG calculations,
one has to combine all degrees of 
freedom for one energy shell to one single site. This
results in an exponentially growing local Fock space when increasing the
number of conduction bands. Therefore, NRG is currently limited to a
very few
conduction bands. As different
conduction bands only couple directly at the impurity, it is easy to
split these bands in a DMRG calculation. It is even possible to
split the spin-up and spin-down electron degrees of freedom of the
conduction bands, as 
also they only couple at the impurity. 

\begin{figure}[tb]
\includegraphics[width=0.8\linewidth]{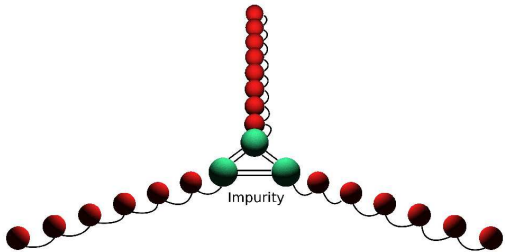}
\caption{(Color online) Structure of the discretized three-site impurity
  model. The impurity sites (green) are the only sites including a
  two-particle interaction, coupling different conduction bands.  As
  in the SIAM, the conduction bands are non-interacting sites with
  nearest neighbor hopping.}\label{3imp}  
\end{figure}

For testing the abilities of DMRG to calculate spectral functions for
multi-impurity models, we have performed calculations for a
three-site impurity model, as visualized in Fig. \ref{3imp}.
We include only an intra-band density-density interaction $U$ on the
impurity site and a hopping $t^\prime$, coupling different impurity
sites. The Hamiltonian thus reads 
\begin{eqnarray}
H&=&H_{Imp}+H_{cond}\\
H_{Imp}&=&\epsilon_f\sum_{l\sigma}f_{l\sigma}^\dagger f_{l\sigma}+ U\sum_{l}f_{l\uparrow}^\dagger f_{l\uparrow}
f_{l\downarrow}^\dagger f_{l\downarrow}\nonumber\\
&&+t^\prime\sum_{l\neq
  n\sigma}f_{l\sigma}^\dagger f_{n\sigma}\nonumber\\
H_{cond}&=&V\sum_{l\sigma} \left(f_\sigma^\dagger
c_{l0\sigma}+c_{l0\sigma}^\dagger
f_{l\sigma}\right)\nonumber\\
&&+\sum_{li\sigma}t_i\left(c_{li\sigma}^\dagger
c_{li\sigma}+c_{li\sigma}^\dagger c_{li\sigma}\right)\nonumber,
\end{eqnarray}
in which $l$, $n$ are channel indices, running from $1$
to $3$, $i$ is the site-index
in a conduction band, and $\sigma$ the spin-index.
The shown impurity Hamiltonian is just an
example. In principle it is no problem to include other types of
interactions on the impurity.

We start the DMRG calculation in one of the conduction electron chains
including only the corresponding impurity site, firstly neglecting
the other two chains. Having initialized this chain, all matrices are
copied to the other chains. 
Thus, all three chains are
described by the same basis set, just changing the band index.
Copying the basis sets will prevent breaking the symmetries between
the chains during the initialization, but is only
justified in the case in which all conduction bands are supposed to
be equal.
The
next DMRG step is the most time-consuming step. All three chains,
which are described by $m$ states, have to be coupled as a
Y-junction.\cite{guo2006} Thus,  
the dimension of the Fock space is now $m^3$. 
The diagonalization of the Hamiltonian is not the
only time-consuming 
part, also the diagonalization of the density matrix, becomes very
time-consuming, as it has dimension $m^2$ and must be 
diagonalized completely. 
After combining two conduction bands, there is one block describing
the basis of two chains.
Having calculated this ``two-chain'' block, DMRG works in the usual way
improving the basis set of the remaining chain, always calculating the
ground state for the whole system. 
To accelerate the convergence we always copy the improved basis set to
the other two chains. 

\begin{figure}[tb]
\includegraphics[width=0.8\linewidth]{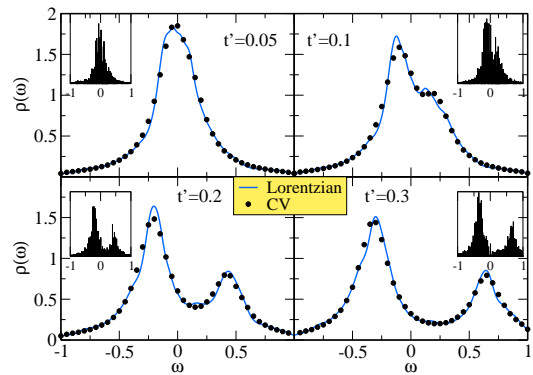}
\caption{(Color online) Spectral function for a non-interacting
  three-Impurity Anderson Model, $\Delta=0.1$, $U=0$, $\epsilon_f=0$. The impurities are coupled by a hopping
  $t^\prime=\{0.05,0.1,0.2,0.3\}$ to each other. We compare the direct correction
  vector result (CV), $\eta=0.05$, to the Lorentzian broadening of the
  calculated  peak structure. The DMRG used $m=300$ kept states.}\label{3impU0}  
\end{figure}

To calculate the peak structure as described in the previous section, we
now have to use a two-block configuration, because the impurity is not
located at an open boundary anymore. The used configuration consists
of a left-block, describing two chains, and a right-block, describing
the other chain. For the purpose of complete
  diagonalization for calculating the spectral function, the blocks
  are additionally truncated to $100$ states. Results for 
the non-interacting case are given in Fig. \ref{3impU0}. 
For all the shown multi-impurity calculations we use a
linear discretization consisting of 50 sites for each channel.
We compare the calculated correction vector result, the
calculated peak structure, and a Lorentzian broadening of the peaks
for four different hopping amplitudes between the
impurities. Calculating the exact result for this hopping Hamiltonian
and using the same Lorentzian-broadening as in the correction vector,
agrees with the calculated correction vector points. 
The physics can already be understood by coupling only three sites to a
triangle by a hopping $t^\prime$. This three-site Hamiltonian has two
single-particle-excitation, namely at $-t^\prime$ and
$2t^\prime$. Coupling a conduction 
band to each of the impurities, results in a broadening of each of
those excitations. Furthermore, by coupling three sites to a triangle,
the particle-hole-symmetry is immediately broken.
\begin{figure}[tb]
\includegraphics[width=0.8\linewidth]{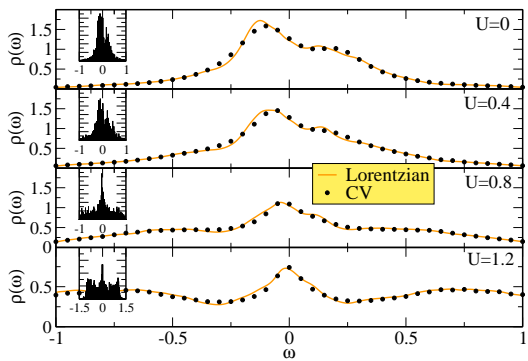}
\caption{(Color online) Spectral function for a three-Impurity Anderson Model,
  $\Delta=0.1$. The impurities are coupled by a hopping 
  $t^\prime=0.1$ to each other. The panels show different interaction
  strengths $U=(0.0; 0.4; 0.8; 1.2)$ ($\epsilon_f=-U/2$) on the impurity sites.
We show the calculated correction vector result (CV) and the
Lorentzian broadening of the peak structure, which can be seen in the
inset. The DMRG used $m=300$ kept states.
\label{3impU}}
\end{figure}

Figure \ref{3impU} shows the results for the interacting case with
$\Delta=0.1$. The hopping
between the impurities is chosen as $t^\prime=0.1$, resulting in peaks at
$\omega=-0.1$ and $\omega=0.2$ for the non-interacting model. Switching
on the interaction, one can observe, how an asymmetric
Kondo-resonance appears by merging of the two non-interacting peaks.
Comparing between the correction vector points and the Lorentzian
broadening of the calculated peak structure, we find again
good agreement. So even in this difficult model, for which the
impurity is not located anymore at the boundary, we are able to
calculate a reliable peak structure.

An interaction or
hopping between different impurities introduces entanglement between
the chains. Combining the 
chains to a single block is therefore the step, at which the truncated weight
can be very high, as the dimension is reduced from $m^2$ to $m$. 
\begin{figure}[tb]
\includegraphics[width=1\linewidth]{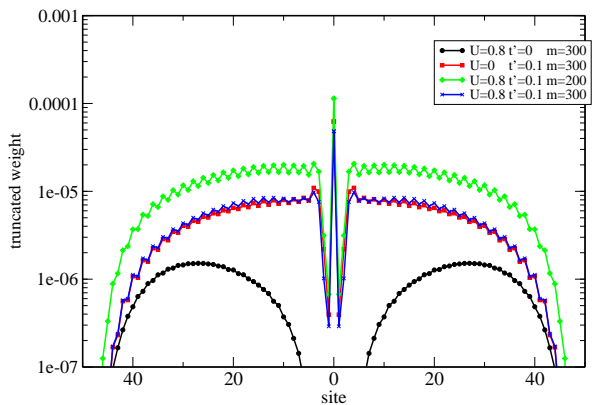}
\caption{(Color online) Truncated weight for a ground state
  calculation of the three-impurity system ($\Delta=0.1$,
  $\epsilon_f=-U/2$). As reference a calculation without coupling
  between 
  the impurities is shown ($t^\prime=0$). The label ``site''
  corresponds to the middle block of the used three-block DMRG
  algorithm. ``site=0'' represents the point at which two chains are
  combined to one block.
 (Further explanation in text.)
\label{3imptrunc}}
\end{figure}
Figure \ref{3imptrunc} shows three examples for the truncated weight
for such three-impurity calculations. For comparison the truncated
weight for a calculation without coupling between the impurities is
shown, too. The index ``site=0'' represents the point at which two chains are
  combined to one block.
Without any coupling between the impurities, the ground
state can be written as product state of three
single-impurity models, for which the impurity is located at an open
boundary. Therefore, the truncated weight vanishes at
``site=0'' in this case. When the impurities are
coupled, combining two chains to one block produces a significant
peak in the truncated weight. Keeping $m=200$ states, the truncated weight
when combining the chains, is approximately $\delta\rho\approx
10^{-4}$. Increasing the kept states to $m=300$, decreases the truncated
weight to $\delta\rho\approx 5\cdot 10^{-5}$. Keeping $m=300$ states
in the shown examples, the ground state energy
as well as impurity expectation values like occupation do not change in
their first 5 digits during DMRG sweeping after convergence is
achieved. When calculating the spectral function and thus including
the correction vector into the density matrix, 
the truncated weight is increased.
However, the calculated spectral functions for the
non-interacting case ($U=0$) agree very well with the exact spectral
functions, as already mentioned above.

It is in principal easy to go even beyond three conduction bands. 
Increasing the number of conduction bands will
eventually make it impossible to couple all bands at the same DMRG
step. However, it is always possible to iteratively couple the different
conduction bands, which, of course, will even increase the truncation
error. Nevertheless, we have demonstrated that it is possible to
calculate multi-impurity properties using DMRG, making it possible to
use DMRG in multi-orbital DMFT or cluster-DMFT calculations.

\section{Conclusions}
We have used the DMRG to calculate spectral functions for single and
multi-impurity models. We discretized the conduction band electrons
using the same scheme as in NRG. However, using DMRG gives more freedom of
choosing the intervals for the discretization, because DMRG does not rely
on energy separation. Therefore, the hopping parameters in
the discretized conduction electron chain can be arbitrary, and
orbital and spin degrees of freedom of the conduction electrons can be
split. A disadvantage of DMRG arises when calculating spectral
functions, as the DMRG basis is very optimized for the ground
state. Nevertheless, there are different methods to calculate spectral functions
within DMRG, but usually the resolution is limited. This limitation can
lead to unphysical solutions when using DMRG as an impurity solver in DMFT.

We here used the correction vector method, from which a
Lorentzian convoluted spectral function can be calculated. 
To improve the resolution, 
we used 
a complete diagonalization
of the DMRG basis in special configurations and calculated a
peak structure of the spectral function. 
It is essential to adapt the
DMRG basis via the correction vector so that also excitation are
well described. By performing these diagonalization for different
correction vectors, a dense set of peaks can be calculated.
It should be noted that the additional time cost is negligible to the rest
of the correction vector calculation.
This peak structure can be directly analyzed as for example
extracting the weight and position of sharp features. It also helps
understanding what structures can actually be resolved, as it exactly
shows the basis from which the correction vector spectral function is
calculated. 
Besides this, these peaks can be convoluted using an arbitrary
broadening function creating a smooth spectral function. Thus, it is
easy to change the convolution from Lorentzian (correction vector) to
any other function.
Finally, this peak structure represents a
very good approximation to a deconvolution of the correction
vector spectral function. 

In the second part of this article, we used this technique for a
model, in which three Anderson impurity models are coupled via a
hopping. We show, that it is again possible to use 
the introduced algorithm to
obtain a peak structure of the spectral function.
We thus are able to calculate precise spectral function for this
model, making future application of DMRG as Cluster-DMFT solver possible.

\begin{acknowledgments}
RP wants to thank Norio Kawakami, Thomas Pruschke, Masaki Tezuka, and
Piet Dargel
for helpful comments and discussions.
RP thanks the Japan Society for the Promotion of Science (JSPS)
together with the Alexander von Humboldt-Foundation
for a postdoctoral fellowship. Parts of the calculations were performed
at ISSP (Tokyo).
\end{acknowledgments}

\end{document}